\documentstyle[12pt,aasms4,graphics]{article}
\newcommand{\etal}{et~al.~}

\begin{document}
\oddsidemargin=0mm

\title{
The Katzman Automatic Imaging Telescope Gamma-Ray Burst Alert System,
and Observations of GRB 020813 
}

\author{Weidong Li, Alexei V. Filippenko,
Ryan Chornock, Saurabh Jha \\
Email: (wli, alex, rchornock,sjha)@astro.berkeley.edu}

\affil{Department of Astronomy, University of California, Berkeley,
CA 94720-3411} 

\slugcomment{Submitted to PASP}

\begin{abstract}

We present the technical details of the gamma-ray burst (GRB) alert system of
the Katzman Automatic Imaging Telescope (KAIT) at Lick Observatory, and the
successful observations of the GRB 020813 optical afterglow with this system.
KAIT responds to GRB alerts robotically, interrupts its pre-arranged program,
and takes a sequence of images for each GRB alert.  A grid-imaging procedure is
used to increase the efficiency of the early-time observations. Different
sequences of images have been developed for different types of GRB alerts. With
relatively fast telescope slew and CCD readout speed, KAIT can typically
complete the first observation within 60~s after receiving a GRB alert,
reaching a limiting magnitude of $\sim 19$.

Our reduction of the GRB 020813 data taken with KAIT shows that unfiltered
magnitudes can be reliably transformed to a standard passband with a precision
of $\sim$5\%, given that the color of the object is known. The GRB 020813
optical afterglow has an exceptionally slow early-time power-law decay index,
though other light-curve parameters and the optical spectral index are fairly
typical of GRBs.

\end{abstract}


\section{INTRODUCTION}

Although the study of gamma-ray bursts (GRBs) has been revolutionized in the
past few years because of the successful operation of several space GRB
detection experiments [such as the Burst and Transient Source Experiment
(BATSE) on board the Compton Gamma Ray Observatory (CGRO), Meegan \etal 1992;
the BeppoSAX satellite, Boella \etal 1997; and the High Energy Transient
Explorer-2 (HETE-2), Ricker et al. 2001], the origin and nature of GRBs remain
enigmatic. Much of the difficulty in studying GRBs results from their short
duration ($\sim$ 1--100~s), the generally poor precision of their reported
positions, and the rapidly fading brightness of their counterparts in the
lower-energy passbands (X-ray, optical, and radio). Since the detection and
observation in these lower-energy passbands during and shortly after a GRB
potentially holds the key to significant progress in understanding the central
engine of GRBs and could provide valuable clues to their progenitors
(M\'esz\'aros 2001), it is critical to have nearly real-time position
determination and prompt follow-up observations in all passbands shortly after
the trigger of a GRB.

HETE-2 is currently the only GRB detector capable of localizing and
disseminating GRB positions in nearly real-time, and the GRB Coordinates
Network (GCN; Barthelmy \etal 1994) provides a link between GRB triggers and
observers/telescopes, including real-time distribution of GRB alerts to a
handful of sites with robotic equipment. The successful observation of the GRB
990123 optical afterglow (OA) by the Robotic Optical Transient Search
Experiment (ROTSE; Akerlof \& McKay 1999; Akerlof \etal 1999) only 22~s after
the GRB trigger, and observations by the Livermore Optical Transient Imaging
System (LOTIS) and Super-LOTIS (Park \etal 2001), demonstrate that it is useful
to have robotic telescopes that can respond to GRB alerts automatically.

In this paper we describe the GRB alert system with the 0.76~m (30 inch)
Katzman Automatic Imaging Telescope (KAIT), currently the world's largest
robotic telescope capable of responding to GRB alerts and doing real-time GRB
optical follow-ups. KAIT can reach a limiting magnitude of $\sim$19 in a 20~s
unfiltered exposure, and can slew at a relatively fast speed (5$^\circ$
s$^{-1}$).  The main goal of the KAIT GRB alert system is not to search widely
for OAs of GRBs, as KAIT's ability to detect a particular GRB OA is limited by
its small field of view ($6\farcm7 \times 6\farcm7$), but rather to obtain a
high-quality early-time light curve of a GRB OA once it is detected in
the KAIT
field. With these early-time data we will be able to constrain properties of
GRBs such as (1) the temporal behavior of the OA emission, (2) breaks and
undulations in the light curves, and (3) other derived parameters such as the
initial Lorentz factor of the blast wave, the jet opening angle, and the
ambient medium density.  The relatively deep, rapid observations made by KAIT
will also provide useful constraints on some of the well-localized ``dark
GRBs," in which no optical afterglow is seen.  The best example to
date of KAIT's abilities was provided by KAIT observations of GRB
021211 (Li et al. 2003), which showed evidence of reverse shock
emission at early times.

The paper is organized as follows. Section 2 contains a description of KAIT,
the GRB alert system, and the performance of the system in its first year of
operation. Section 3 reports the successful observations of the GRB 020813 OA
with this system, and the transformation of the observed unfiltered photometry
to a standard system. Conclusions are drawn in Section 4.

\section{THE KAIT GRB ALERT SYSTEM}

\subsection{KAIT}

KAIT is the third robotic telescope in the Berkeley Automatic Imaging Telescope
(BAIT) program. The technical details, the operation concept, and the
scientific objectives of the earlier BAIT systems can be found in Filippenko
(1992), Richmond, Treffers, \& Filippenko (1993), and Treffers et al. (1995),
while those of KAIT specifically are in Li \etal (2000), Filippenko \etal
(2001), and Filippenko \etal (2003, in preparation). Here we briefly summarize
the operation concept of KAIT and the results from the ongoing supernova (SN)
search.

KAIT consists of a 0.76~m diameter primary with a Ritchey-Chreti\'en mirror
set. The telescope has a focal ratio of $f/8.2$ and a plate scale of
$33\farcs2$ mm$^{-1}$ at the focal plane. An off-axis autoguider enables long
exposure times. The thermoelectrically cooled CCD camera is an Apogee AP7b with
a SITe $512 \times 512$ pixel back-illuminated chip (pixel size 24~$\mu$m),
which has a pixel scale of 0$\farcs$8. We use only the central $500 \times 500$
pixels in observations due to some chip defects near the edges, yielding a
total field of view (FOV) of $6\farcm7 \times 6\farcm7$. KAIT has a filter
wheel with 20 slots, including a set of standard Johnson $UBV$ and Cousins $RI$
filters.  A weather station monitors the outside and telescope temperatures,
humidity, wind speed, rain, and cloud cover, and it sends signals to close the
dome slit whenever conditions are hazardous to the telescope system.

All of the hardware is controlled by computers. The whole system starts
automatically each day in the afternoon. The hardware status is checked and
initialized, and the observations for the night are scheduled according to all
the active request files. The bias, dark current, and twilight flatfield images
are automatically taken, and observations of the arranged targets begin when
the Sun is $8^\circ$ below the horizon. A focusing routine is run every 90 min
to ensure good focus.  During bad weather, such as high winds, high humidity,
fog, rain, or completely overcast skies, the slit is closed automatically, and
the system takes a ``nap" for 10 min. It tries to do observations again after
the nap, and keeps trying every 10 min thereafter. At the end of the night the
system is shut down automatically; it then goes to ``sleep" until waking up
again in the afternoon.

The primary science project carried out with KAIT is the Lick Observatory
Supernova Search (LOSS; e.g., Treffers \etal 1997; Filippenko \etal 2001),
which recently became part of the Lick Observatory and Tenagra Observatory
Supernova Searches (LOTOSS; e.g., Schwartz \etal 2000). Depending on the
season, about 5,000--10,000 nearby galaxies are monitored every 2 to 15 days by
LOSS; over the course of the year, the sample size is about 14,000
galaxies. Images are automatically processed and candidate supernovae (SNe) are
flagged; these are subsequently examined by a group of research assistants
(most of whom are undergraduate students). Promising SN candidates are then
reobserved, and the confirmed SNe are reported to the Central Bureau of
Astronomical Telegrams, where the International Astronomical Union Circulars
are issued.

LOSS is the world's most successful search engine for nearby SNe, discovering
20 in 1998, 40 in 1999, 36 in 2000, 68 in 2001, and 82 in 2002. Moreover, most
of the LOSS SNe were discovered while young and thus were especially suitable
for detailed photometric and spectroscopic studies. The large number of SN
discoveries is also ideal for statistical studies such as the SN rate in
galaxies of different Hubble types; see van den Bergh, Li, \& Filippenko (2002)
for the first step of this process, morphological classification of the host
galaxies.  For more details on the publications utilizing LOSS data, see the
web site http://astron.berkeley.edu/$\sim$bait/kait.html and Filippenko \etal
(2003, in preparation).

\subsection{The GRB OA Observation Program}

Because KAIT does all observations robotically, many years ago we considered
incorporating a GRB OA observation program into the system, but the idea was
not implemented until the launch of HETE-2.  Before HETE-2, the only prompt
localizations available were from BATSE; with large positional uncertainties
($\sim 1^\circ - 10^\circ$; Briggs et al. 1999), however, it was impractical to
implement a real-time strategy with KAIT (FOV only $6\arcmin.7 \times
6\arcmin.7$).  With the launch of HETE-2, which is capable of providing nearly
real-time GRB localization through GCN alerts and significantly improved GRB
position uncertainty (about 20$\arcmin$--40$\arcmin$, sometimes better than
10$\arcmin$), the chance of detecting GRB OAs in the small FOV of KAIT has
dramatically improved. We thus designed a GRB OA observation program in late
2001, and the system has been operating since the beginning of the year 2002.

The GRB OA observation program connects to the GCN through socket
communications, the fastest way (nearly real-time) to receive GRB alerts.  Upon
receiving a GRB alert, the program responds differently depending on the
contents of the alert. If there is no information on the GRB position, or the
alert indicates that the trigger was not due to a GRB, only an email message is
sent to several team members. When there are GRB coordinates in the alert, and
there is no indication that the trigger was not due to a GRB, the program
checks the possibility of observing the GRB field, including the following:

\begin{enumerate}

\item{The position uncertainty. If the uncertainty exceeds $1^\circ$, the field
will not be observed.}

\item{The angular distance from the Moon. Depending on the phase of the Moon,
GRB fields within different distances (e.g., 45$^\circ$ during full moon and
25$^\circ$ during quarter moon) from the Moon are not observed.}

\item{The time of the trigger. If the altitude of the Sun is $<8^\circ$ below
the horizon, it is considered twilight or daytime and the field will not be
observed.}

\item{The telescope limit. Fields that have declination south of $-29^\circ$ or
north of $+70^\circ$, or absolute hour angle exceeding 4.5 hr, are not observed
due to the mechanical limits of KAIT. As HETE-2 GRBs are preferentially
anti-solar, those that are triggered during nighttime hours in California are
likely to be within (or close to) the KAIT mechanical limits, except during the
summer months.}

\end{enumerate}

If the field is determined to be not observable for any reason, an email
message is sent to the team members, and the telescope resumes its normal
program (usually the SN search), of course continuing to be alert for the next
GRB trigger. If, on the other hand, the field can be observed right away, the
program immediately terminates the on-going observation by KAIT, and interrupts
the previous robotically scheduled observing process.\footnote{Currently, KAIT
cannot stop an exposure in progress, so the shutter remains open during the
move to the GRB position. This may even result in a double exposure, of two
different parts of the sky, as was the case for GRB 021211 (Li et al.  2003).}
The program then takes a sequence of images of the GRB field, after which the
normal automatic observing process resumes.

Thus far, we have chosen to conduct the GRB observations without filters (i.e.,
in the unfiltered mode). The advantage of this is that deeper limiting
magnitudes can be achieved in relatively shorter exposures: the limiting
magnitude of a 20~s unfiltered exposure is 19.0 (3$\sigma$) under favorable
observing conditions, while that of a 300~s exposure is 21.5. However, there
are two disadvantages to this choice. First, it is difficult to compare
unfiltered magnitudes from different telescopes --- but as we show in \S 3,
unfiltered photometry can be transferred to the standard Cousins $R$ band with
a relatively large color term, and the resulting $R$ magnitudes have
uncertainties of about 5\%. Second, the unfiltered observation lacks the
critical color information to place physical constraints on GRB models --- but
observations of GRB afterglows are still at a stage where only a few very early
(within ten minutes of the burst) light curves have been obtained, and our
current goal is to expand the sample with more high-quality early unfiltered
light curves. We plan to switch to observing with standard filters in the
future.

Our original sequence of observations (before 2002 Nov. 22; UT dates are used
throughout this paper) for each GRB field was arranged as follows. First, five
$3 \times 20$~s grid images (discussed below) were taken, then two 60~s
exposures, one 120~s guided exposure, one 300~s guided exposure, and finally,
four 120~s guided exposures that combine to make a $2 \times 2$ mosaic image
centered on the GRB position, covering a region about $12\farcm7 \times
12\farcm7$.  The whole sequence took about half an hour to observe.  The
sequence was arranged so that relatively short exposures were observed at the
beginning when the GRB OA is expected to be bright, and longer exposures were
observed at the end when the OA is expected to become more faint.  The $2
\times 2$ mosaic observations increased the coverage of the GRB field, but no
attempt was made to cover the whole error box of each GRB.

The ROTSE observations of GRB 990123 (Akerlof \etal~1999) have the earliest
detection of a GRB OA, and indicate that the early-time light curve of
an OA can vary on short timescales: the OA of GRB 990123 brightened by 3 mag in
25 s, then declined by 1 mag in another 25 s.  It is important to have good
temporal coverage to study these short timescale variations. For this reason,
we have developed a procedure to obtain grid images of a GRB field without
reading out the CCD.  When the first 20~s exposure is finished, the telescope
is offset to the east by $0\farcm6$ (45 pixels), and the CCD shutter is closed
but the CCD is not read out. This repeats twice to produce a $3 \times 20$~s
grid, and the whole image is finally read out.  Although the readout time for
the KAIT CCD camera is relatively short (10~s), by saving two readouts the
observation efficiency is increased by about 20\%, which is significant for the
time-critical early OA observations.

The grid procedure adds difficulties to the identification and photometry of
individual stars in the final image.  However, the offset between the
individual 20~s exposures (45 pixels) is big enough to not complicate the
photometric reductions too much, and the normal images obtained after the grid
procedures offer references to identify individual stars in the grid.  There is
also the probability that the GRB OA will be superimposed on another star in
one or more of the dithering positions, but since the grid images are intended
to target bright GRB OAs, the point-spread-function (PSF) fitting procedure in
the photometry packages helps determine the individual brightness of the star
and the OA. Figure 1 shows a grid image of the GRB 020813 field taken by the
KAIT GRB observation program.  The GRB OA is marginally detected in each 20~s
exposure (see \S 3 for more details on the observations of GRB 020813). For
comparison, Figure 2 shows a non-grid image of the GRB 020813 field.

After the sequence of images of a GRB field is observed, an email message is
sent to the team members.\footnote{Recently, we modified the system so that a
team member is woken up, when there is a particularly promising GRB alert.}
Those who are aware of the observations will then check the data, identify
possible OAs and alert the GRB community, and arrange further observations. If
no real-time analysis is done, the images are checked the next day by one or
more of the team members.

The requested sequence of images has recently been modified, as we learn from
our previous attempts to observe GRB OAs. The observations of GRB 020813 showed
that the grid images at the beginning of the sequence were not deep enough to
detect the GRB OA, and since most GCN alerts with initial positions of GRBs are
later than one hour after the bursts, we changed the sequence to start with a
relatively shallow (10~s) and a relatively deep (40~s) exposure, followed by a
set of grid images. This latter sequence was followed for GRB 021211, our most
successful set of observations of a GRB OA (see next section for details), which
demonstrates that grid images are, in fact, effective when GRB positions are
determined in real time by HETE-2. Our current (April 2003) sequence thus {\it
starts} with the grid images (first $3 \times 5$~s, then $3 \times 10$~s and $3
\times 20$~s), and continues with longer unguided (60~s) and guided (120~s and
300~s) exposures. We expect the sequence to continue to evolve, reflecting our
strategy for observing GRB OAs.

The GRB alert program has also been modified so that incoming GCN notices are
studied to determine whether they are first-time alerts, or follow-up alerts of
a previous GRB. In the latter case, a different sequence starting with long
exposures commences, since the GRB OA is likely to have already faded
considerably.

\subsection{The System Performance}

The KAIT GRB alert program has been in operation since late-December 2001. Our
statistics show that during the year 2002, a successful connection was made
between KAIT and GCN only 70\% of the time. The cause of the connection
failures is unknown, but is probably related to network problems. The
connectivity has improved to over 90\% of the time since November 2002.

Table 1 shows some HETE-2 alerts that KAIT received, but failed to conduct
real-time observations for various reasons in 2002. Apparently, most of the
failures were caused by bursts occurring during the daytime at KAIT.  The only
solution to this problem is to have a network of robotic telescopes at
different longitudes.  Other reasons for failures were bad weather, the
connection problem mentioned above, and bursts occurring near or beyond the
mechanical or software limits of KAIT. In particular, the HETE-2 on-board
localization was derived for GRB 021004 almost immediately, and the GCN alert
arrived at KAIT when the GRB was at a western hour angle of 4.4 hr, which is
within the mechanical limit of KAIT (4.6 hr), but the software had limited the
observations to hour angles within $\pm$4 hr. We subsequently changed the
software limit to $\pm$4.5 hr, in order to maximize the area of sky potentially
accessible to KAIT.

Table 2 shows the HETE-2 alerts that KAIT received and responded to in real
time. For the discussions in this paper, $t$ = 0 is defined as the GRB on-board
trigger time. The position of GRB 020813 was released in a GCN notice at $t$ =
4.2 min when it was still daytime at KAIT. We commenced remote manual
observations of the GRB field when it was dark ($t = 103.7$ min), but this was
interrupted by the KAIT GRB alert program when it received an updated notice at
$t$ = 112.7 min.  KAIT finished a sequence of images of the GRB 020813 field,
and then resumed its regular observing schedule. A third GCN notice was
received at $t$ = 184.3 min, and KAIT responded with the same sequence of
images.  Using robotic images from the Palomar 48-inch telescope, Fox, Blake,
\& Price (2002) discovered the OA of GRB 020813.  Our manual and automatic
observations had successfully imaged this OA position, but not all images are
deep enough to detect the OA.  We also remotely (but manually) obtained images
of the GRB 020813 field the next night. Preliminary results from these
observations were reported by Li, Filippenko, \& Chornock (2002) and by Li,
Chornock, \& Filippenko (2002); a detailed analysis is presented in \S 3.

The GCN alerts with positions for GRB 021104, GRB 021112, and GRB 021113 were
received when KAIT was conducting its regular SN search, and the GRB alert
program successfully responded with sequences of images. Unfortunately, no OA
was discovered for these GRBs from our images, or from images obtained
elsewhere.

GRB 021211 is KAIT's most successfully observed GRB OA. The position of GRB
021211 was relayed by HETE-2 at $t$ = 22~s.  KAIT responded to the GCN notice
at $t$ = 32~s, and began to slew across much of the sky to the GRB position. It
finished pointing at $t$ = 108~s (but the dome slit began to clear the
telescope at $t$ = 105~s), with 48~s left in a 600~s follow-up exposure of SN
2002he. Because currently we cannot stop an exposure in progress, the image
ended up with the GRB 021211 field superimposed on the SN 2002he field (but
nevertheless a useful image which provided the earliest KAIT measurement of GRB
021211). KAIT then took a sequence of images, starting with a 10~s exposure and
continuing with exposures of 40~s, 3$\times$20~s grid, 60~s, 120~s (guided),
and 300~s (guided). Three additional GCN alerts with updated positions for GRB
021211 were received during the night, and KAIT responded with the same
sequence of images. The position of the OA of GRB 021211 (Fox \& Price 2002)
was successfully observed on most images. Preliminary results from these
observations were reported by Li et al. (2002) and Chornock et al. (2002), and
a detailed analysis is presented elsewhere (Li et al. 2003). Our observations
of GRB 021211 not only provide one of the fastest detections of a GRB OA, but
also show evidence for a reverse-shock component in the early optical emission
and a break at $t \approx 10$ min. Our photometry of GRB 021211, which has high
temporal resolution, also suggests that either the object underwent a dramatic
color change at early times, or that there are small-scale variations
superimposed on the power-law decay of the reverse-shock emission (Li et
al. 2003).

The experience we gained from the observations of the GRB 021211 OA prompted us 
to
make improvements to the existing GRB alert program. For example, the response
procedure was simplified to speed up the observations, the sequence of images
was changed to start with short $3 \times 5$~s grid images, and follow-up GCN
alerts are observed with a sequence starting with relatively long exposures
instead of short ones.

Table 2 also shows that KAIT can typically start the first observation at about
20--40~s after receiving the GCN alert. With our improved procedure, tests show
that KAIT can start the first exposure as early as 8~s after receiving a
position from the GCN notices.  Table 2 also shows that with small-number
statistics, KAIT is more successful with GRBs that have near real-time HETE-2
localizations (e.g., GRB 020813 and GRB 021211) than with GRBs that are
localized more than 600 s after the bursts.

\section{OBSERVATIONS OF GRB 020813} 

\subsection{Data Reduction}

GRB 020813 was detected by HETE-2 at 2:44 on 2002 August 13 (Villasenor et al.
2002). The gamma-ray light curve shows a series of pulses that last $>$ 125~s,
typical of a bright, long-duration burst. The flight localization was reported
in a GCN notice at $t$ = 4.2 min. The OA of GRB 020813 was discovered by Fox,
Blake, \& Price (2002) in images taken at $t \approx$ 2 hr with the Palomar
1.2~m Oschin Schmidt telescope; independent analysis of our own images also
revealed the OA (Li, Chornock, \& Filippenko 2002), but we reported the results
later than Fox, Blake, \& Price (2002).  Spectroscopy of the GRB 020813 OA with
the Keck telescope (Price et al. 2002) identifies absorption systems with the
highest redshift measured at $z = 1.254$.  Spectropolarimetric observations
also done with Keck (Barth et al.  2002, 2003) yield a linear polarization of
1.8--2.4\% at $t$ = 4.7--4.9 hr, the first spectropolarimetry ever obtained for
a GRB OA.  A temporal break in the light curve was first suggested by Bloom,
Fox, \& Hunt (2002), and confirmed by us (Li, Chornock, \& Filippenko 2002),
Malesani et al. (2002), and Gladders \& Hall (2002b).

We obtained 25 images of the GRB 020813 field on 2002 August 13, among which 9
were manually and 16 were automatically observed. A total of 45 individual
exposures of the GRB 020813 OA were taken due to our grid procedure.  An
additional 6 images were manually taken on 2002 August 14. Because of the
faintness and the rapid decline of the OA, not all of our images are useful.

Since our observations were obtained without filters, it is important to
transform the measured magnitudes to a standard photometric system, so that
comparisons with observations taken elsewhere can be made. In reducing
unfiltered data on Type Ia SNe, Riess et al. (1999) demonstrated a
method that treats the unfiltered images as though they were observed through a
very broad filter, and uses color terms to correct for the difference in the
throughput curves, just as we correct the instrumental $BVRI$ magnitudes to the
standard Johnson-Cousins system.  The color terms can be empirically determined
by obtaining $BVRI$ and unfiltered images of Landolt (1992) standard stars
during photometric nights.  Riess et al. found that for a SN~Ia,
whose spectrum has broad absorption and emission features and is far from
stellar, a reliable transformation from unfiltered magnitudes to a standard
passband can be achieved at a precision of $\sim$5\%.  Since most GRB OA
spectra don't have broad absorption and emission features, we expect the
transformation to work even better for the unfiltered GRB OA observations.

We find that the combination of the KAIT optics and the quantum efficiency of
the Apogee CCD camera makes the KAIT unfiltered observations mostly mimic the
$R$ band, so we choose to transform the unfiltered magnitudes to $R$.  The
formula for the differential photometry that transforms the unfiltered
magnitudes to $R$ is as follows:

\begin{equation}
R_{GRB} = (c_{GRB} - c_{LS}) + R_{LS} - CT \times [ (V - R)_{GRB} - (V -
R)_{LS} ], 
\end{equation}

\noindent 
where $R_{GRB}$, $c_{GRB}$, and $(V-R)_{GRB}$ are the $R$ magnitude,
unfiltered magnitude, and $(V - R)$ color of the GRB OA, respectively, while
$R_{LS}$, $c_{LS}$, and $(V - R)_{LS}$ are those of the local standard
star. ``$CT$" is the color term to transform the KAIT unfiltered magnitudes to
$R$, and is found to be $0.27 \pm 0.05$ from several photometric nights.

We have adopted the calibration of the GRB 020813 field by Henden (2002), and
$(V - R)$ = 0.39 mag for GRB 020813 from Gladders \& Hall (2002a).  The final
transformed $R$ magnitudes from the unfiltered KAIT observations are listed in
Table 3.

To cross-check the accuracy of the above transformation, we have independently
measured the extensive $R$-band photometry of the GRB 020813 OA by the 6.5-m
Magellan telescope, which were generously made available to the GRB community
by Gladders \& Hall (2002c). The same calibration of the GRB 020813 field
(Henden 2002) was used. The measured $R$ magnitudes are listed in Table 4.

Figure 3 displays the $R$-band light curve of the GRB 020813 OA from the
unfiltered KAIT observations, while Figure 4 shows that from KAIT, Magellan,
the BAO 1.0-m telescope (Kawabata, Urata, \& Yamaoka 2002), and the TUG 1.5-m
(Kiziloglu et al. 2002) observations.  In these figures we assume no
host-galaxy reddening to GRB 020813 (Barth et al. 2003), and adopt a Galactic
extinction of $A_R = 0.26$ mag (Schlegel, Finkbeiner, \& Davis 1998).  It is
apparent from both figures that there is a break in the light curve of the GRB
020813 OA at $t \approx $ 0.2~d, which is further qualitatively demonstrated by
the change of power-law decay indices before and after the break: $-0.47 \pm
0.04$ at $t$ = 0.0720--0.1363~d from the KAIT observations, and $-1.16 \pm
0.04$ at $t$ = 0.2336--1.1873~d from the KAIT observations ($-1.08 \pm 0.03$
from all the observations shown in Figure 4).  The break appears to be smooth
and gradual. For example, the power-law decay index measured at $t$ =
0.1334--0.2357~d is $-0.81 \pm 0.06$ from the KAIT observations ($-0.83 \pm
0.05$ from all the observations shown in Figure 4), which is in between the
earlier and later power-law decay indices.

To better describe the temporal evolution of the GRB 020813 OA, we fit the
data with a smoothly broken power-law model, which is modified from that of
Beuermann et al. (1999):

\begin{equation}
F_\nu = \frac{ 2^{1/s}\, F_{\nu,b}}{ [\, (t/t_b)^{-\alpha_1s} +
(t/t_b)^{-\alpha_2s}\, ]^{1/s}},
\end{equation} 

\noindent 
where $t_b$ is the time of the break, $F_{\nu,b}$ is the flux at $t_b $, and
$s$ controls the sharpness of the break, with larger $s$ implying a sharper
break. We obtained the following values for the parameters using a
$\chi^2$-minimizing procedure: (1) for the KAIT data only: $t_b =0.14 \pm 0.03$
day, $\alpha_1 = -0.13 \pm 0.05$, $\alpha_2 = -1.21 \pm 0.12$, and $s = 2.2 \pm
0.8$. The reduced $\chi^2$ of the fit is 1.12 for 13 degrees of freedom. (2)
For all the data shown in Figure 4: $t_b = 0.13 \pm 0.03$ day, $\alpha_1 =
-0.21 \pm 0.05$, $\alpha_2 = -1.09 \pm 0.08$, and $s = 3.3 \pm 1.0$. The
reduced $\chi^2$ of the fit is 0.63 for 29 degrees of freedom. The dashed lines
in the upper panels of Figures 3 and 4 show the model fits, while the lower
panels of the two figures show the residuals of the fits.

The reduced $\chi^2$ values of both fits show that eq. (1) is an adequate model
to fit the data. The derived parameters from both fits are consistent with each
other within the uncertainties, indicating that the transformed KAIT $R$ data
are consistent with the other $R$-band measurements. In particular, the superb
Magellan data offer a good consistency check, and it can be seen from both the
model fit (upper panel of Figure 4) and the residual of the fit (lower panel of
Figure 4) that the KAIT data are consistent ($\pm 0.05$ mag) with those of
Magellan, although with much bigger uncertainties.

Our reduction of the KAIT data thus confirms the results of Riess et al.
(1999) that the unfiltered magnitudes can be reliably transformed to a standard
passband with a precision of $\sim$5\%, given that the color term is well
determined and the color of the target is known.  The high throughput of the
unfiltered observations thus provides a unique way to go deep for small to
moderate telescopes, while still providing reasonably accurate photometry in a
passband close to the peak of the unfiltered response function.

However, we need to also emphasize the limitations of this transformation.  An
{\it a priori} requirement for the transformation is that the color of the OA
is known at the time of the unfiltered observations, which is often {\it not}
the case. For example, we assumed that the color of the GRB 020813 OA is
constant at all of the epochs of the KAIT unfiltered observations, and is the
same as reported by Gladders \& Hall (2002a) at times that are only coincident
with the second part of the KAIT observations.  Early GRB OAs are expected to
have a large blue to red evolution (Sari, Piran, \& Narayan 1998), though this
has not yet been confirmed with observations, probably because most of the
filtered observations commenced too late. Clearly, the poor constraints on the
color evolution of GRB OAs is the main impediment to transforming unfiltered
magnitudes to a standard passband. Nevertheless, to expedite comparisons
between unfiltered observations from different observatories, it is important
to apply such a transformation and provide information such as the color term
and the color used in the transformation, so that other observers can correct
the reported photometry with different color estimates. For example, if the $(V
- R)$ color of the GRB 020813 OA is assumed to be $VR$, the photometry reported
in Table 3 will need to be increased by $0.27 \times (0.39 - VR)$ mag.

\subsection{Late-Time Power-Law Decay Index and Spectral Index}

Although the power-law decay index after the break ($\alpha_2$) for the GRB
020813 OA is measured to be $-1.09 \pm 0.08$ from the model fit to all the data
shown in Figure 4 as discussed above, later observations in the optical
(Malesani et al. 2002; Levan et al. 2002) and X-rays (Vanderspek et al. 2002)
indicate a further steepening of the light curve.  This suggests that the GRB
020813 OA has a smoothly changing light curve, as partly demonstrated by the
data shown in Figure 4; it also suggests that our measured value of $\alpha_2$
is only an intermediate result between the early and the late-time
measurements, and that the derived parameters ($t_b$, $\alpha_1$, $\alpha_2$,
$s$) may need to be revised when later data are considered in the model fit.
Unfortunately, such late-time data for GRB 020813 are not available at the time
of writing; thus, we adopt $\alpha_2 = -1.39 \pm 0.05$, which is the average of
values reported by Malesani et al. (2002), Levan et al.  (2002), and Vanderspek
et al. (2002). The power-law decay index before the break ($\alpha_1$) is not
expected to be significantly changed, as it is dominated by the early-time
measurements already presented here, so we adopt $\alpha_1 = -0.21 \pm 0.05$
from the model fit to all the data shown in Figure 4.

The optical spectral index of the GRB 020813 OA was reported as $\beta = -1.06
\pm 0.01$ from red-channel Keck spectra (5600--9400~\AA) taken at $t \approx 6$
hr (Barth et al. 2003).  Levan et al. (2002) reported a shallower optical
spectral index of $\beta = -0.8$ from {\it Hubble Space Telescope} images taken
at $t \approx 98$ hr after the burst.  We adopt the Keck measurement, as it was
made during the time of the photometric observations presented here, and we
correct it for a Galactic reddening of $E(B - V) = 0.11$ mag (Schlegel,
Finkbeiner, \& Davis 1998). The final $\beta$ we use for the GRB 020813 OA is
$-0.73 \pm 0.01$.

\subsection{Discussion}

The current standard model for GRBs and their afterglows is the fireball shock
model (for a review see M\'esz\'aros 2002), in which a fireball expands
relativistically into an ambient medium and decelerates as it sweeps up
matter. The shock between the fireball and the medium accelerates electrons to
relativistic energies, and gives them a power-law differential energy
distribution with index $-p$. The simplest afterglow model (e.g., Sari, Piran,
\& Narayan 1998) predicts that the afterglow flux $f_\nu(t) \propto
t^{\alpha}\nu^\beta$, where $f_\nu(t)$ is the flux at frequency $\nu$, and
$\beta$ is the spectral index of the OA.  Interpretation of the afterglow data
in the framework of these models can, in principle, yield many interesting
parameters of the GRB explosion (e.g., Wijers, Vreeswijk, \& Galama 1999;
Holland et al. 2000) because the properties of the observed light curve of the
afterglow, such as the power-law decay rate $\alpha$, are primarily determined
by the hydrodynamic evolution (e.g., adiabatic or radiative), the geometry
(spherical or collimated), and the density profile of the ambient medium
(constant or stellar-wind type) of the fireball (e.g., Sari, Piran, \& Narayan
1998; Wijers \& Galama 1998; Rhoads 1999; Chevalier \& Li 2000).

A break in the OA light curve is often regarded as a signature of a jet (e.g.,
M\'esz\'aros \& Rees 2000; Kulkarni et al. 1999; Rhoads 1999; Moderski, Sikora,
\& Bulik 2000), either with a fixed opening angle $\theta_0$, or expanding
sideways as well as radially.  A jet with a fixed opening angle $\theta_0$ can
have a break in the light curve when the Lorentz factor decreases to $\sim
1/\theta_0$, due to the geometrical effect that the observer only receives
radiation emitted within the collimated beam.  A jet that expands sideways as
well as radially can have a break in the light curve when the sideways
expansion dominates over the radial expansion.

During the initial phase, however, the GRB outflow looks spherical to the
observer in most current models. Under the assumptions of adiabatic cooling and
an ambient medium of constant density, the predicted power-law decay index in
this phase is $\alpha_1 = - (3p - 3)/4$, when $\nu < \nu_c$ (the cooling
frequency, above which the electrons cool on the dynamical timescale of the
system), and $\alpha_1 = - (3p - 2)/4$ when $\nu > \nu_c$ (e.g., Sari, Piran,
\& Halpern 1999).  The measured $\alpha_1 = -0.21 \pm 0.05$ then implies $p =
1.28$ ($\nu < \nu_c$), or $p = 0.95$ ($\nu < \nu_c$), which would imply a
diverging electron energy. This may be an indication that (1) radiative process
may be significant, though Sari, Piran, \& Narayan (1998) show that a radiative
fireball decays with $\alpha_1 = -4/7$, which is still steeper than the
observations; (2) the ambient medium may not have a constant density --- for an
outflow expanding into a stellar wind environment, Chevalier \& Li (2000)
calculate $\alpha_1 = -1/4$ at early times of a GRB when $\nu > \nu_m$ (the
peak frequency, which corresponds to the minimum-energy electrons), in good
agreement with the observations; (3) energy is being injected into the outflow
even during the afterglow phase (e.g., Zhang \& M\'esz\'aros 2002), a
hypothesis that could be constrained with radio and X-ray observations; or (4)
there is a significant color change for the early afterglow, so that the
transformation from the unfiltered magnitudes to $R$ is incorrect. However, to
achieve a steeper $\alpha_1$, the color of the afterglow will need to change
from red to blue, inconsistent with the model prediction by Sari, Piran, \&
Narayan (1998). Alternatively, there are some theoretical considerations (e.g.,
Bykov \& M\'esz\'aros 1996) to allow the formation of a flat electron spectrum
(i.e., $p < 2$).

GRB 020813 has the slowest pre-break decline rate, $\alpha_1$, of all GRBs with
well measured optical light-curve breaks (which all show $\alpha_1 \lesssim
-0.7$; see, e.g., Figure 3 of Stanek et al. 2001). Afterglow models generally
predict a correlation between the decline rates $\alpha$ and the spectral index
$\beta$; for GRB 020813, however, the observed $\beta = -0.73$ is fairly
typical among GRB OAs. To be more consistent with the models, it is possible to
invoke host-galaxy extinction; as little as $E(B-V) = 0.05$ mag of reddening in
the host would imply an intrinsic $\beta = -0.50 \pm 0.03$, which increases to
$\beta = -0.26 \pm 0.03$ for $E(B-V)_{\rm host} = 0.10$ mag. However, such flat
spectral indices typically imply unreasonable electron indices $p \lesssim 1$,
and thus host-galaxy extinction is unlikely to explain the peculiarities of GRB
020813.

The amplitude of the break ($\Delta \alpha = \alpha_2 - \alpha_1$) is
independent of extinction with the assumption of no color change, and can also
be used to constrain models. For GRB 020813, we find $\Delta\alpha = -1.18 \pm
0.07$, which is also fairly typical among well-observed GRBs. In the framework
of a constant ambient density model, this result favors a sideways-expanding
jet ($\Delta \alpha = \alpha_1/3 - 1 = -1.07 \pm 0.02$) over a jet with a fixed
opening angle ($\Delta \alpha = -3/4$; Rhoads 1999), but as discussed above, a
wind medium may better explain the early decline.

\section{CONCLUSIONS}

In this paper we present the technical details of the GRB alert system at KAIT,
a 0.76~m telescope with relatively high resolution imaging capacity (0$\farcs$8
pixel$^{-1}$). Connected to the GCN via socket communication, KAIT is capable
of responding to GRB alerts in real time, and of capturing a pre-arranged
sequence of images automatically. The observations are done in the unfiltered
mode to reach deeper limiting magnitudes in relatively shorter exposures. A
grid image procedure is used to increase the efficiency of the early-time
observations. The sequence of images starts with short-exposure grid images,
and continues with increasingly longer exposures to compensate for the decline
of a GRB OA.  A different sequence that starts with long exposures is followed
in response to a GCN notice with updated positions for a previous GRB.  All of
these efforts are aimed to getting good early-time photometry rather than
searching wide regions for the OAs of GRBs.

Our reduction of the GRB 020813 data taken with KAIT shows that unfiltered
magnitudes can be reliably transformed to a standard passband with a precision
of $\sim$5\%, if the color of the object is known.

The available data on GRB 020813 from KAIT, Magellan, and elsewhere (as
reported in the GCN Circulars) indicate that the GRB 020813 OA has an
exceptionally slow early-time power-law decay index, though other light-curve
parameters and the optical spectral index are fairly typical. GRB 020813 thus
presents an interesting challenge to afterglow models for GRBs; additional
observations (for instance, in the radio or X-rays) may shed light on the
situation for this particular burst. Continued automatic observations with KAIT
of other bursts in the future will help determine whether GRB 020813 remains
exceptional in a larger sample.

\acknowledgments

A.V.F. is grateful to Stan Woosley for encouraging him, years ago, to use KAIT
for the optical follow-up of GRBs having sufficiently precise coordinates.  We
thank Mike Gladders and Pat Hall for making their superb Magellan data
available and for assistance with the reductions, and Scott Barthelmy for his
help setting up the KAIT GRB alert system.  We also thank Shiho Kobayashi and
Bing Zhang for useful discussions.  The work of A.V.F.'s group at
U. C. Berkeley is supported by National Science Foundation grant AST-9987438,
as well as by the Sylvia and Jim Katzman Foundation. KAIT was made possible by
generous donations from Sun Microsystems, Inc., the Hewlett-Packard Company,
AutoScope Corporation, Lick Observatory, the National Science Foundation, the
University of California, and the Katzman Foundation. S.J. thanks the Miller
Institute for Basic Research in Science at U. C. Berkeley for support through a
research fellowship.

\renewcommand{\arraystretch}{0.75} 

\begin{deluxetable}{cccrrrrcl}
\tablecaption{KAIT Responses to HETE-2 Alerts in 2002; No Real-Time
Observations\tablenotemark{a}}
\tablehead{
\colhead{Trig.}&\colhead{UT date}&
\colhead{UT time}&\colhead{$\Delta t$\tablenotemark{b}}&
\colhead{$\alpha_{J2000}$}&\colhead{$\delta_{J2000}$}& 
\colhead{Err.\tablenotemark{c}}& \colhead{OA\tablenotemark{d}} &
\colhead{Reason\tablenotemark{e}}\\ 
\colhead{}&\colhead{(yy/mm/dd)}&\colhead{(hh:mm:ss)}&\colhead{(min)}&\colhead{(d
eg.)}
&\colhead{(deg.)}&&&}
\startdata
1896& 02/01/24& 10:41:15.13&  85.6& 143.206& $-$11.460& 24$\arcmin$& yes&bad 
weather    \\
1902& 02/01/27& 20:57:24.73& 105.6& 123.774& +36.742& 24$\arcmin$& no &daytime   
    \\
1939& 02/03/05& 11:55:25.05& 596.5& 190.762& $-$14.552& 50$\arcmin$& yes&daytime 
      \\
1959& 02/03/17& 18:15:31.42& 533.6& 155.837& +12.744& 36$\arcmin$& no &bad 
weather    \\
1963& 02/03/31& 16:32:28.76&  40.4& 199.142& $-$17.875& 20$\arcmin$& yes&daytime 
      \\
2042& 02/05/31& 00:26:18.71&  88.0& 228.688& $-$19.360& 77$\arcmin$& no 
&daytime\tablenotemark{f}       \\
2081& 02/06/25& 11:25:49.32& 173.6& 311.060& +07.170& 28$\arcmin$& no &daytime   
    \\
2257& 02/08/12& 10:41:43.98&   9.1& 309.699& $-$05.393& 28$\arcmin$& no &no 
connection\\
2262& 02/08/13& 02:44:19.16&   6.5& 296.658& $-$19.588&  2$\arcmin$& 
yes&daytime\tablenotemark{g}       \\
2275& 02/08/19& 14:57:35.82&  98.1& 351.849&  +6.269&  4$\arcmin$& no &daytime   
    \\
2380& 02/10/04& 12:06:13.57&   0.8&   6.737& +18.929&  4$\arcmin$& yes&telescope 
limit\tablenotemark{h} \\
2397& 02/10/16& 10:29:00.74& 104.2&   2.768& +49.139&659$\arcmin$& no &telescope 
limit \\
\enddata
\tablenotetext{a}{Only alerts with position information are listed.}
\tablenotetext{b}{The time delay (minutes) between the GRB trigger and the
first reported GCN position.} 
\tablenotetext{c}{The uncertainty of the given position (diameter).}
\tablenotetext{d}{Whether an OA was detected by us or by others.}
\tablenotetext{e}{The reason why no KAIT real-time observations were made.}
\tablenotetext{f}{The GRB was manually observed later, and the results were
reported by Li, Chornock, \& Filippenko (2002).}
\tablenotetext{g}{The GRB was manually and automatically observed later. See 
text for
details.}
\tablenotetext{h}{The GRB was initially within KAIT's mechanical limit, but not
within its software limit; see text for details.}
\end{deluxetable}

\begin{deluxetable}{cccrrrrrccc}
\tablecaption{KAIT Responses to HETE-2 alerts in 2002;
Real-Time Observations}
\tablehead{
\colhead{Trig.}&\colhead{UT date}&
\colhead{UT time}&\colhead{$\Delta 
t$\tablenotemark{a}}&\colhead{Obs(t)\tablenotemark{b}}&
\colhead{$\alpha_{J2000}$}&\colhead{$\delta_{J2000}$}& 
\colhead{Err.}& \colhead{OA} &
\colhead{t1\tablenotemark{c}} &\colhead{Exp\tablenotemark{d}} \\ 
&\colhead{(yy/mm/dd)}&\colhead{(hh:mm:ss)}&\colhead{(min)}&\colhead{(min)}&
\colhead{(deg.)}
&\colhead{(deg.)}&&&\colhead{(s)}&\colhead{(s)}}
\startdata
2262& 02/08/13& 02:44:19.16&   4.2&112.7& 296.658& $-$19.588&  2$\arcmin$& yes& 
143 &3$\times$20 \\
2262& 02/08/13& 02:44:19.16&   4.2&184.3& 296.658& $-$19.588&  2$\arcmin$& yes& 
63 &3$\times$20 \\
2434& 02/11/04& 07:01:02.93& 166.0&166.0&  58.452& +37.953& 52$\arcmin$& no & 24 
&3$\times$20\\
2448& 02/11/12& 03:28:15.89&  81.2& 81.2&  39.127& +48.849& 40$\arcmin$& no & 
108 &3$\times$20\\
2448& 02/11/12& 03:28:15.89&  81.2&121.1&  39.127& +48.849& 40$\arcmin$& no & 28 
&3$\times$20\\
2449& 02/11/13& 06:38:56.90& 120.9&120.9&  23.473& +40.462& 27$\arcmin$& no & 40 
&3$\times$20\\
2449& 02/11/13& 06:38:56.90& 120.9&153.6&  23.473& +40.462& 27$\arcmin$& no & 31 
&3$\times$20\\
2493& 02/12/11& 11:18:34.03&   0.4&  0.5& 122.228&  +6.735& 28$\arcmin$& yes& 
108\tablenotemark{e}& 10\\
2493& 02/12/11& 11:18:34.03&   0.4& 34.9& 122.228&  +6.735& 28$\arcmin$& yes& 35 
 & 10\\
2493& 02/12/11& 11:18:34.03&   0.4& 66.8& 122.228&  +6.735& 28$\arcmin$& yes& 23 
 & 10\\
2493& 02/12/11& 11:18:34.03&   0.4&131.0& 122.250&  +6.739&  4$\arcmin$& yes& 45 
 & 10\\
\enddata
\tablenotetext{a}{The time delay (minutes) between the GRB trigger and the
first reported GCN position.} 
\tablenotetext{b}{The time (minutes after the GRB trigger) of the alert message 
that 
activates KAIT.}
\tablenotetext{c}{The response time when KAIT started the first observation.}
\tablenotetext{d}{The exposure time for the first observation.}
\tablenotetext{e}{This image is a superposition of two fields; see text for
details. The exposure actually started at $t = 105$~s, when the dome slit began
to clear the telescope.}
\end{deluxetable}

\begin{deluxetable}{lccccc}
\tablecaption{KAIT Photometry of the GRB 020813 OA}
\tablehead{
\colhead{UT time\tablenotemark{a}}&\colhead{$t$ (hr)\tablenotemark{b}}&
\colhead{Exp. (s)}&\colhead{$R$\tablenotemark{c}}&
\colhead{$\sigma_R$\tablenotemark{c}} & \colhead{Obs.\tablenotemark{d}}
}
\startdata 
4:28:00 &  1.728 &  30 & 17.90 & 0.05 & m \\
4:28:46 &  1.740 &  30 & 17.98 & 0.06 & m \\
4:29:32 &  1.754 &  30 & 17.81 & 0.05 & m \\
4:32:12 &  1.797 & 120 & 17.91 & 0.04 & m \\
4:36:43 &  1.874 & 300 & 17.94 & 0.04 & m \\
4:46:11 &  2.030 &  30 & 17.92 & 0.06 & a \\
4:46:58 &  2.045 &  30 & 17.95 & 0.07 & a\\
4:50:21 &  2.100 & 300 & 18.00 & 0.07 & a\\
5:56:28 &  3.201 &  30 & 18.16 & 0.06 & a\\
5:57:14 &  3.216 &  30 & 18.08 & 0.05 & a\\
6:00:36 &  3.271 & 300 & 18.23 & 0.02 & a\\
7:35:18 &  4.850 & 120 & 18.58 & 0.04 & m\\
7:39:15 &  4.915 & 120 & 18.54 & 0.05 & m\\
8:20:41 &  5.606 & 120 & 18.73 & 0.08 & m\\
8:23:45 &  5.657 & 300 & 18.67 & 0.05 & m\\
5:02:55 & 26.311 & 600 & 20.59 & 0.18 & m\\
7:00:58 & 28.277 & 600 & 20.77 & 0.23 & m\\
\enddata
\tablenotetext{a}{UT time at the middle of the exposure. All are on
2002 August 13, except the last two which were on August 14.}
\tablenotetext{b}{Starting time after the GRB, in hours.}
\tablenotetext{c}{Magnitude and $1\sigma$ uncertainty derived from unfiltered 
observations; see text for details.}
\tablenotetext{d}{``Observer" of the images: those marked with ``m'' 
were manually obtained via remote observations;
while those marked with ``a'' were obtained
by the automatic GRB alert program.}
\end{deluxetable} 

\begin{deluxetable}{lccc}
\tablecaption{Magellan Photometry of the GRB 020813 OA\tablenotemark{a}}
\tablehead{
\colhead{UT time\tablenotemark{b}}&\colhead{$t$ 
(hr)\tablenotemark{c}}&\colhead{$R$\tablenotemark{d}}&
\colhead{$\sigma_R$\tablenotemark{d}}
}
\startdata 
6:41:52 &  3.967 &  18.40 & 0.03 \\
6:54:14 &  4.173 &  18.43 & 0.02 \\
6:56:02 &  4.205 &  18.44 & 0.02 \\
7:08:22 &  4.409 &  18.48 & 0.03 \\
7:09:52 &  4.435 &  18.49 & 0.02 \\
7:11:22 &  4.459 &  18.49 & 0.02 \\
7:12:52 &  4.483 &  18.50 & 0.02 \\
7:18:50 &  4.584 &  18.52 & 0.03 \\
7:23:03 &  4.653 &  18.54 & 0.03 \\
7:30:00 &  4.769 &  18.56 & 0.03 \\
1:14:11 & 22.507 &  20.30 & 0.03 \\
1:15:41 & 22.531 &  20.31 & 0.03 \\
1:17:11 & 22.555 &  20.32 & 0.03 \\
1:18:41 & 22.581 &  20.33 & 0.03 \\
\enddata
\tablenotetext{a}{Measured from the publically available data on
site ftp.ociw.edu under pub/gladders/ GRB/GRB020813. All exposures were
of 60~s duration.}
\tablenotetext{b}{UT time at the middle of the exposure. All are on
2002 August 13, except the last four which were on August 14.}
\tablenotetext{c}{Starting time after the GRB, in hours.}
\tablenotetext{d}{$R$-band magnitude and $1\sigma$ uncertainty.}
\end{deluxetable}

\newpage

\begin{figure}
{\plotfiddle{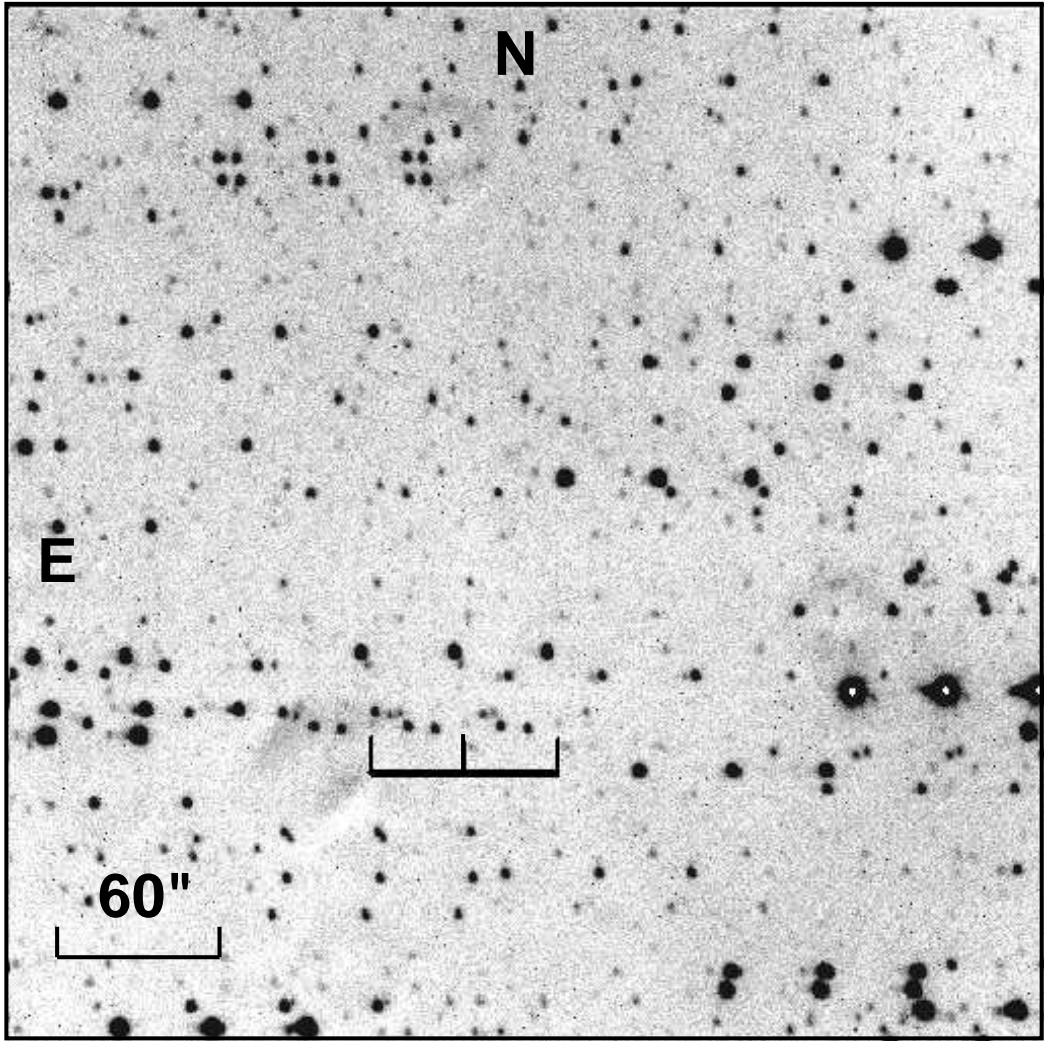}{6.5in}{0}{85}{85}{-8}{-50}}
\figcaption{An unfiltered KAIT grid image of the GRB 020813 field. 
The OA of GRB 020813, which is quite faint, is marked at three positions. }
\end{figure}

\begin{figure}
{\plotfiddle{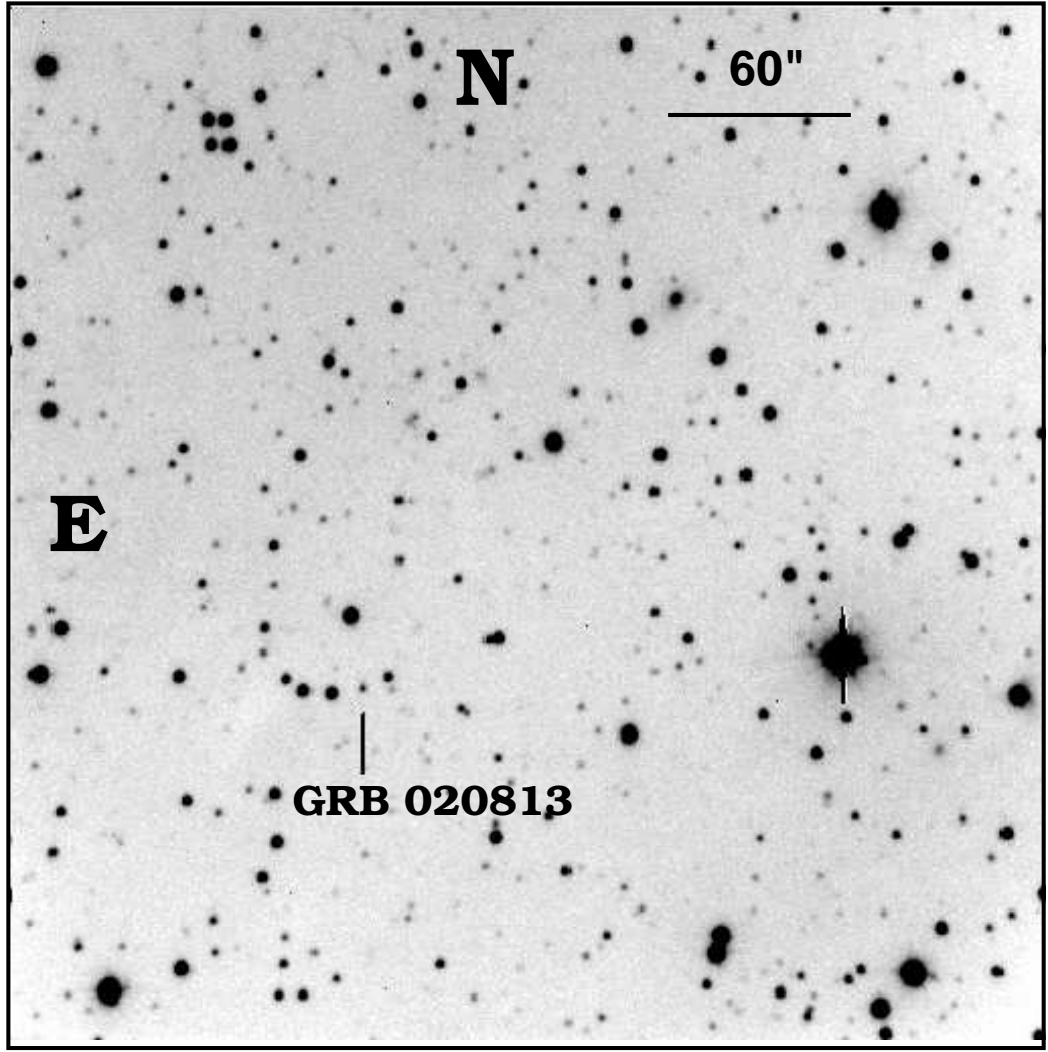}{6.5in}{0}{85}{85}{-8}{-50}}
\figcaption{An unfiltered KAIT non-grid image of the GRB 020813 field. The OA 
of GRB 020813 is marked.}
\end{figure}

\begin{figure}
{\plotfiddle{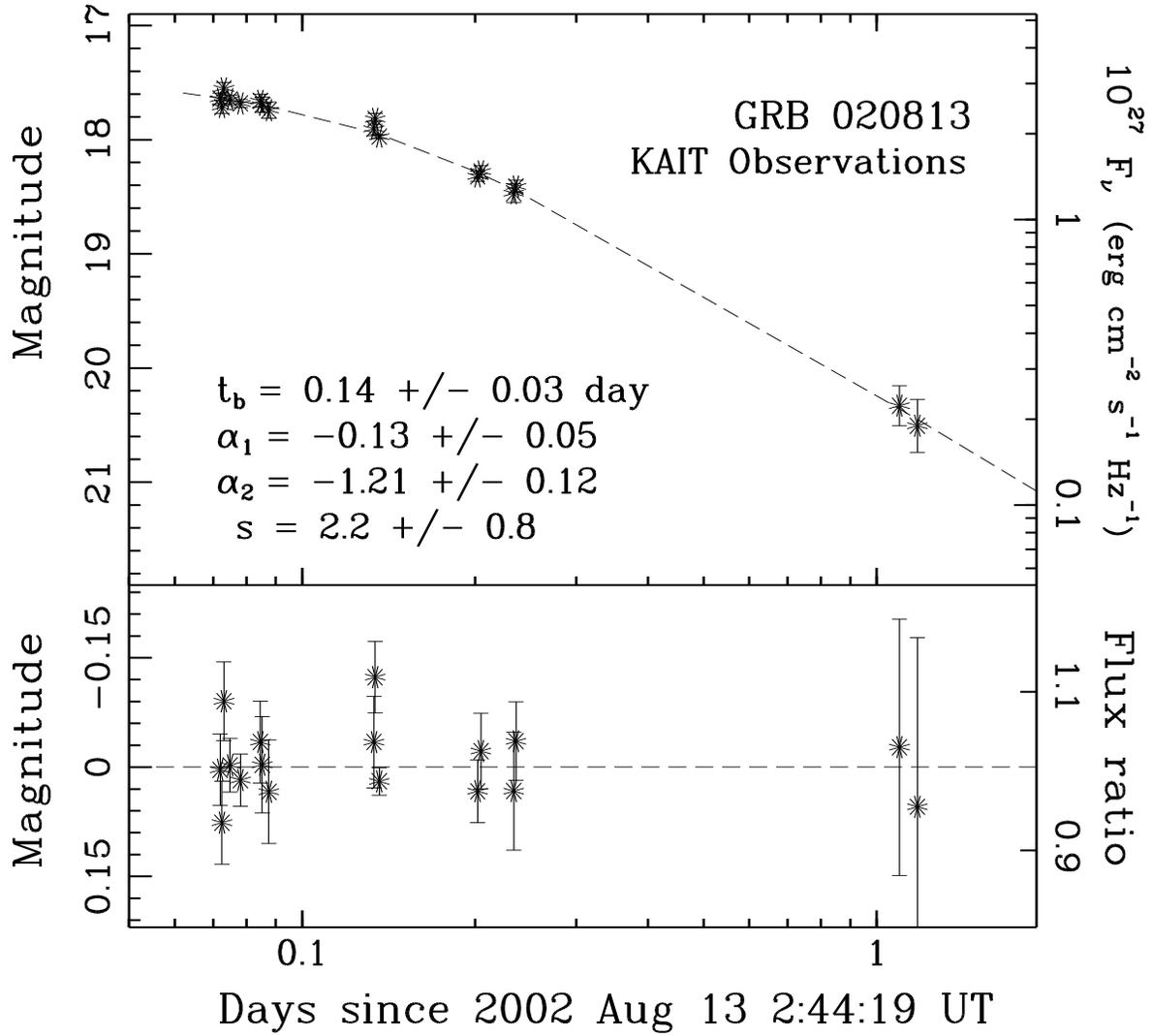}{5.8in}{-90}{80}{80}{-090}{480}}
\figcaption{Light curve of the GRB 020813 OA from the KAIT observations.
The upper panel shows the broken power-law fit, while the lower panel shows 
the residuals of the fit. }
\end{figure}

\begin{figure}
{\plotfiddle{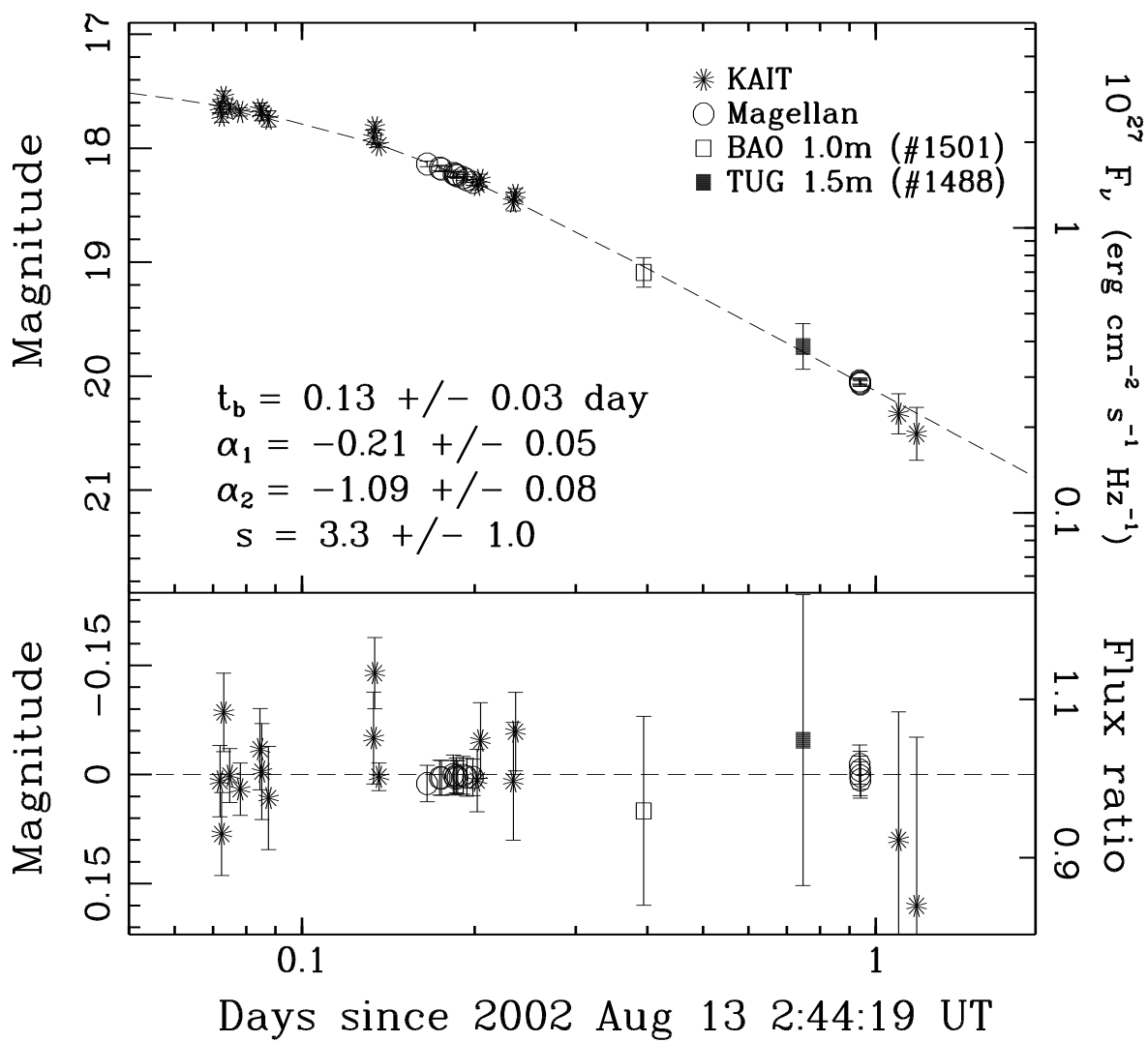}{5.8in}{-90}{80}{80}{-090}{480}}
\figcaption{Light curve of the GRB 020813 OA from various sources. The upper
panel shows the broken power-law fit, while the lower panel shows the 
residuals of the fit. }
\end{figure}

\end{document}